\documentclass[conference]{IEEEtran}
\usepackage{times}
\usepackage{amsmath}
\usepackage{amssymb}
\usepackage{epsfig, latexsym,subfigure, cite}

\newcommand{\eqnlabel}[1]{\label{eqn:#1}}
\newcommand{\eqnref}[1]{(\ref{eqn:#1})}

\providecommand{\xv}{\mathbf{x}}
\providecommand{\yv}{\mathbf{y}}
\providecommand{\uv}{\mathbf{u}}
\providecommand{\vv}{\mathbf{v}}

\providecommand{\Uv}{\mathbf{U}}
\providecommand{\Vv}{\mathbf{V}}
\providecommand{\Xv}{\mathbf{X}}
\providecommand{\Yv}{\mathbf{Y}}

\providecommand{\Xc}{{\cal X}}
\providecommand{\Yc}{{\cal Y}}
\providecommand{\Wc}{{\cal W}}

\providecommand{\Cc}{{\cal C}}

\providecommand{\Nc}{{\cal N}}

\providecommand{\Ycc}{{\tilde \Yv}_1^{i+1}}
\newcommand{\A}{A^{(n)}_{\epsilon}}

\newtheorem{MyTheorem}{Theorem}

\newtheorem{MyCorollary}{Corollary}

\newtheorem{MyLemma}{Lemma}

\begin{document}
\title{The Discrete Memoryless Multiple Access Channel with Confidential Messages}

\author{\authorblockN{Ruoheng Liu,
Ivana Mari\'c,
Roy D. Yates and
Predrag Spasojevi\'c}
\authorblockA{WINLAB, Rutgers University\\
Email: \{liurh,ivanam,ryates,spasojev\}@winlab.rutgers.edu}
}

\maketitle
\footnotetext[1]{This work was supported by NSF Grant NSF ANI 0338805.}

\begin{abstract}
A multiple-access channel is considered in which messages from one encoder are confidential.
Confidential messages are to be transmitted with perfect secrecy, as measured by equivocation at
the other encoder. The upper bounds and the achievable rates for this communication situation are
determined.
\end{abstract}
\section{Introduction}
%
We consider a two-user discrete multiple-access channel in which one user wishes to communicate
confidential messages to a common receiver while the other user is permitted to eavesdrop. We
refer to this channel as the {\em multiple access channel with confidential messages} (MACC) and
denote it $(\Xc_1 \times \Xc_2, p(y,y_1|x_1, x_2), \Yc \times \Yc_1)$.  The communications system
is shown in Figure~\ref{f:Figure1}. The ignorance of the other user is measured by equivocation.
This approach was introduced by Wyner \cite{Wyner75} for the wiretap channel, a scenario in which
a single source-destination communication is eavesdropped. Under the assumption that the channel
to the wire-tapper is a degraded version of that to the receiver, Wyner determined the
capacity-secrecy tradeoff. This result was generalized by Csisz{\'{a}r and K{\"{o}rner who
determined the capacity region of the broadcast channel with confidential messages
\cite{CsiszarKorner78}. The Gaussian wire-tap channel was considered in~\cite{CheongHellman78}.

In this paper, we determine the bounds on the capacity region of the MACC, under the requirement
that the eavesdropping user is kept in total ignorance. The results characterize the rate penalty
when compared to the conventional MAC \cite{Liao72, Ahlswede71} due to the requirement that one
message is kept secret.

It is apparent from the results that eavesdropping by user $1$ will hurt the achievable rate of
user $2$. As illustrated in the last section by an example in which the half-duplex constraint is
imposed, the eavesdropper should give up on listening all together, thus maximizing rates of both
users. The moral of the example is that either user $1$ will make both himself and the other user
miserable by eavesdropping more and thus reducing both its own and other user's ability to
transmit; or, it will make both of them happy if it decides not to listen.
We note that, although user $2$ cannot know exact times when user $1$ is eavesdropping, it is
enough for user $2$ to know the eavesdropping probability (or equivalently, the fraction of time
 user $1$ is listening), to adjust its code  rate accordingly.  This information can be
considered public, since it is known to the common receiver.

\section{Channel model and statement of result} \label{s: Channel}

A discrete memoryless MAC with confidential messages consists of finite sets $\Xc_1, \Xc_2,\Yc,
\Yc_1$ and a conditional probability distribution $p(y,y_1|x_1, x_2)$.  Symbols $(x_1,x_2) \in
{\Xc}_1 \times {\Xc}_2$ are channel inputs and $(y,y_1) \in {\Yc} \times {\Yc}_1$ are channel
outputs at the receiver and encoder $1$, respectively. The channel $p(y|x_1,x_2)$ is a MAC
channel, and the channel $p(y,y_1|x_1, x_2)$ is a wire-tap channel. Each encoder $t$, $t=1,2$,
wishes to send an independent message $W_t \in \{1, \ldots, M_t\}$ to a common receiver in $n$
channel uses. The channel is memoryless and time-invariant in the sense that
\begin{equation} \eqnlabel{channel}
p(y_{1,i},y_{2,i}|\xv^i_1,\xv^i_2,\yv^{i-1}_1,\yv^{i-1}_2)=p(y_{1,i},y_{2,i}|x_{1,i},x_{2,i})
\end{equation}
where $\xv_t^i=\begin{bmatrix} x_{t,1},& \ldots, &x_{t,i} \end{bmatrix}$.
To simplify notation, we drop the superscript when $i=n$.
\begin{figure}[t]
 \centerline{\includegraphics[width=0.9\linewidth,draft=false]{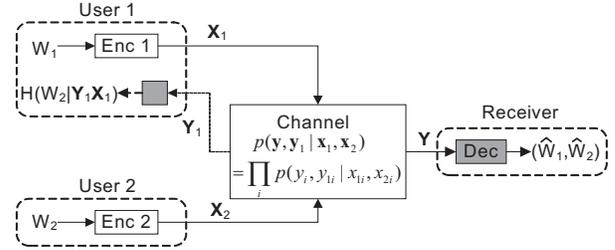}}
  \caption{\small System Model}
  \label{f:Figure1}
\end{figure}
A deterministic encoder $g$  for user $1$ is a mapping $g: \Wc_1 \rightarrow \Xc_1^n$ generating
codewords
\begin{equation}\eqnlabel{codewordsx1}
\xv_1=g(w_1).
\end{equation}
A stochastic encoder $f$ for user $2$ is
specified by a matrix of conditional probabilities $f(\xv_2|w_2)$, where
$\xv_2\in\mathcal X^{n}_2$, $w_2\in{\mathcal W_2}$,
is the private message set, and
\[\sum_{\xv_2}f(\xv_2|w_2)=1.\]
Note that $f(\xv_2|w_2)$ is the probability that the message $w_2$ is
encoded as channel input $\xv_2$.

The decoding function is given by a mapping $\phi: {\mathcal Y}^{n}\rightarrow \Wc_1 \times
\Wc_2$.

The implicit assumption in our model is that user $1$ observes the sequence $\Yv_1$ in block
fashion. This prevents user $1$ from using symbols $Y_1$ for encoding its own messages, as reflected
in the encoding function~\eqnref{codewordsx1}.
This restriction of our model is made for the sole purpose of making the problem easier to solve and understand.

An $(M_1,M_2,n,P_e)$ code for the channel consists of two encoding functions $f,g$, decoding
function $\phi$ such that the {\it average probability of error} of the code is
\begin{equation}
\begin{split}
P_{e}&=\sum_{(w_1,w_2)}  \frac{1}{M_1 M_2} P\{\phi(\yv)\neq (w_1,w_2)|(w_1,w_2)~\text{sent}\}
\end{split}
\end{equation}
The level of ignorance of user $1$ with respect to the confidential message is
measured by the normalized equivocation $(1/n)H(W_2|\Xv_1,\Yv_1)$.

A rate pair $(R_1, R_2)$ is achievable for the MACC
if, for any $\epsilon > 0$, there exists a $(M_1, M_2, n, P_e)$ code such that
\begin{equation}
M_t \ge 2^{nR_t} \hspace{0.1in} t=1,2, \hspace{0.1in} P_{e} \le \epsilon \label{Hlim}
\end{equation}
\begin{equation} \eqnlabel{Hlim2}
  R_2-\frac{1}{n}H(W_2|\Xv_1,\Yv_1) \le \epsilon.
\end{equation}

The capacity region of the MACC is the closure of the set of all achievable rate pairs $(R_1,R_2)$.

The next two theorems show the outer bound and the achievable rates and are the main results of
this paper.

Let $\Cc_U$ be a closure of the union of all $(R_1,R_2)$ satisfying
\begin{align}
R_1& \le I(X_1;Y|X_2) \nonumber\\
R_2& \le I(V; Y|U,X_1)-I(V;Y_1|U,X_1) \nonumber\\
R_1+R_2& \le I(X_1,V; Y)-I(V;Y_1|U,X_1)  \label{ub3}
\end{align}
for some joint distribution
\begin{align}
~~~&p(u,v,x_1,x_2,y,y_1)\nonumber\\
&= p(u)p(v|u)p(x_1|u)p(x_2|v)p(y,y_1|x_1,x_2) \eqnlabel{jpdf}
\end{align}
where $U$ and $V$ are auxiliary random variables satisfying $U\rightarrow V \rightarrow (X_1,X_2)
\rightarrow (Y,Y_1)$.
\begin{MyTheorem}  {\it (Outer Bound)} \label{upperbound}
For any achievable rate pair $(R_1,R_2)$ in MACC it holds that $(R_1,R_2) \in \Cc_U.$
\end{MyTheorem}
\begin{MyTheorem} \label{maintheorem} {\it (Achievability)}
The rates in the closure of the union of all $(R_1, R_2)$ satisfying
\begin{align}
R_1& \le I(X_1;Y|U,V) \nonumber \\
R_2& \le I(V; Y|U,X_1)-I(V;Y_1|U,X_1)\nonumber\\
R_1+R_2& \le I(X_1,V; Y|U)-I(V;Y_1|U,X_1)  \eqnlabel{R1R2}
\end{align}
for a joint distribution $p(u,v,x_1,x_2,y,y_1)$ that factors as \eqnref{jpdf}.
\end{MyTheorem}

\section{Outer Bound}
\begin{proof}  {\it (Theorem~\ref{upperbound})}

We next show that any achievable rate pair satisfies
\begin{align}
R_1& \le I(X_1;Y|X_2,Q)\\
R_2& \le I(V; Y|U,X_1,Q)-I(V;Y_1|U,X_1,Q)   \eqnlabel{R2QOUTER}\\
R_1+R_2& \le I(U,X_1,V; Y|Q)-I(V;Y_1|U,X_1,Q)  \eqnlabel{upperboundQ}
\end{align}
for some product distribution $U \rightarrow V \rightarrow (X_1,X_2) \rightarrow (Y,Y_1)$ that
factor as \eqnref{jpdf} and an independent timesharing random variable $Q$. Then, the approach of
\cite[Thm. $14.3.3$]{Cover} and the observation that Markovity  $U \rightarrow V \rightarrow
(X_1,X_2) \rightarrow Y$ implies  $U \rightarrow (V,X_1) \rightarrow Y$, will prove the claim.


Consider a code $(M_1,M_2,n,P_e)$ for the MACC.
Applying Fano's inequality results in
\begin{equation}
H(W_1,W_2|\Yv) \le P_e \log(M_1M_2-1)+h(P_{e}) \triangleq n\delta_n
\end{equation}
where $\delta_n \rightarrow 0$ as $P_e \rightarrow 0$.
It follows that
\begin{equation}\eqnlabel{Fano1C}
H(W_1,W_2|\Yv) =H(W_1|\Yv)+H(W_2|\Yv,W_1)\le  n\delta_n
\end{equation}

We first consider the bound on $R_1$.
\begin{align}\eqnlabel{R1}
nR_1
&=H(W_1) \nonumber \\
&=I(W_1;\Yv)+H(W_1|\Yv) \nonumber \\
&\le^{(a)}I(W_1;\Yv)+n\delta_n  \nonumber \\
&\le^{(b)} I(\Xv_1(W_1);\Yv)+n\delta_n \nonumber \\
&\le^{(c)} I(\Xv_1;\Yv|\Xv_2)+n\delta_n \nonumber \\
&=^{(d)} \sum_{i=1}^n H(Y_i|\Xv_2, \Yv^{i-1})-\sum_{i=1}^nH(Y_i|\Yv^{i-1},\Xv_1,\Xv_2) \nonumber \\
&\hspace{0.3in}+n\delta_n \nonumber\\
&\le^{(e)} \sum_{i=1}^n H(Y_i|X_{2i})-\sum_{i=1}^nH(Y_i|X_{1i},X_{2i})+n\delta_n \nonumber \\
&= \sum_{i=1}^n I(X_{1,i};Y_i|X_{2,i})+n\delta_n
\end{align}
where $(a)$ follows from from Fano's inequality \eqnref{Fano1C}; $(b)$ from \eqnref{codewordsx1};
$(c)$ from the independence of $\Xv_1,\Xv_2$; $(d)$ from the chain rule; $(e)$ from the fact that
the conditioning decreases entropy and from the memoryless property of the channel
\eqnref{channel}.

Following the approach in \cite[Sec. $14.3.4$]{Cover}, we introduce a uniformly distributed random
variable $Q, Q \in \{1, \ldots , n \}$. Equation \eqnref{R1} becomes
\begin{equation}\eqnlabel{R1Q}
\begin{split}
nR_1 & \le \sum_{i=1}^n I(X_{1,i};Y_i|X_{2,i})+n\delta_n\\
& = \sum_{i=1}^n I(X_{1,i};Y_i|X_{2,i}, Q=i)+n\delta_n\\
&= nI(X_{1,Q};Y_Q|X_{2,Q},Q)+n\delta_n\\
&= nI(X_1;Y|X_2,Q)+n\delta_n
\end{split}
\end{equation}
where $X_1=X_{1,Q}, X_2=X_{2,Q}, Y=Y_Q$. Distributions of new variables depend on $Q$ in the same way as the distributions of $X_{1,i}, X_{2,i}, Y_i$ depend on $i$.

Next, we derive the bound on $R_2$. Note that the perfect security \eqnref{Hlim2} implies
\begin{equation} \eqnlabel{P17}
nR_2-n\epsilon \le H(W_2|\Xv_1,\Yv_1).
\end{equation}
Hence, we consider the bound on $H(W_2|\Xv_1, \Yv_1)$.
\begin{align}\eqnlabel{equivocation}
H(W_2|&\Xv_1,\Yv_1) \nonumber \\
& = H(W_2|\Xv_1)-I(W_2;\Yv_1|\Xv_1) \nonumber \\
&= I(W_2;\Yv|\Xv_1)+H(W_2|\Yv,\Xv_1) -I(W_2;\Yv_1|\Xv_1)\nonumber \\
& \le I(W_2;\Yv|\Xv_1) -I(W_2;\Yv_1|\Xv_1)+n\delta_n
\end{align}
where the inequality follows from Fano's inequality \eqnref{Fano1C}. We next use a similar
approach as in \cite[Sect.V]{CsiszarKorner78} to bound equivocation $H(W_2|\Xv_1,\Yv_1)$ in
\eqnref{equivocation}.

We  denote $\Ycc=[Y_{1,i+1}, \ldots, Y_{1,n}]$ and use the chain rule to obtain
\begin{align}
I(W_2&; \Yv|\Xv_1)\nonumber\\
&= \sum_{i=1}^n I(W_2;Y_i| \Yv^{i-1}, \Xv_1)\nonumber\\
&=\sum_{i=1}^n I(W_2,; Y_i| \Ycc, \Yv^{i-1}, \Xv_1) + \Sigma_1-\Sigma_2\eqnlabel{P1}\\
I(W_2&; \Yv_1|\Xv_1) \nonumber\\
&= \sum_{i=1}^n I(W_2;Y_{1i}| \Ycc, \Xv_1) \nonumber\\
&= \sum_{i=1}^n I(W_2,; Y_{1,i}| \Ycc, \Yv^{i-1}, \Xv_1)+ {\hat \Sigma_1}-{\hat \Sigma_2}
 \eqnlabel{P2}
\end{align}
where
\begin{align*}
\Sigma_1&= \sum_{i=1}^n I(\Ycc; Y_i| \Yv^{i-1}, \Xv_1) \\ 
\Sigma_2&= \sum_{i=1}^n I(\Ycc; Y_i| \Yv^{i-1}, \Xv_1, W_2) \\ 
{\hat \Sigma}_1&= \sum_{i=1}^n I(\Yv^{i-1}; Y_{1,i}| \Ycc, \Xv_1) \\ 
{\hat \Sigma_2}&= \sum_{i=1}^n I(\Yv^{i-1}; Y_{1,i}| \Ycc, \Xv_1,W_2). 
\end{align*}
\begin{MyLemma} \label{lem1}
$\Sigma_1={\hat \Sigma_1}$ and  $\Sigma_2={\hat \Sigma_2}.$
\newline
\begin{proof}
 Proof follows the approach in \cite[Lemma $7$]{CsiszarKorner78}.
\end{proof}
\end{MyLemma}
We let
\begin{equation} \eqnlabel{U}
U_i=(\Yv^{i-1} \Ycc \Xv^{i-1}_1 {\tilde \Xv}^{i+1}_1)\\
\end{equation}
\begin{equation}
V_i=(W_2,U_i) \eqnlabel{V}
\end{equation}
 in \eqnref{P1} and \eqnref{P2} and obtain respectively
\begin{equation} \eqnlabel{P7}
I(W_2; \Yv|\Xv_1)= \sum_{i=1}^n I(V_i; Y_i|U_i, X_{1,i})+ \Sigma_1-\Sigma_2
\end{equation}
\begin{equation} \eqnlabel{P8}
I(W_2; \Yv_1|\Xv_1)= \sum_{i=1}^n I(V_i; Y_{1,i}| U_i, X_{1,i})+ {\hat \Sigma_1}-{\hat \Sigma_2}
\end{equation}
We follow the same approach as in \eqnref{R1Q} to obtain
\begin{align} \eqnlabel{P9}
\frac{1}{n}\sum_{i=1}^n I(V_i; Y_i|U_i, X_{1,i})
&=\frac{1}{n}\sum_{i=1}^n I(V_i; Y_i|U_i, X_{1,i}, Q=i) \nonumber \\
&= I(V_Q; Y_Q|U_Q, X_{1,Q}, Q) \nonumber \\
&= I(V; Y|U, X_1,Q)
\end{align}
where $V=V_Q, Y=Y_Q, X_1=X_{1,Q}, U=U_Q$.
Similarly,
\begin{equation} \eqnlabel{P10}
\frac{1}{n}\sum_{i=1}^n I(V_i; Y_{1,i}| U_i, X_{1,i})= I(V; Y_1| U, X_1,Q)
\end{equation}
where $Y_1=Y_{1,Q}$. From the memoryless property of the channel \eqnref{channel}, it follows that
$V \rightarrow (X_1,X_2) \rightarrow (Y, Y_1).$


Using \eqnref{P9} in \eqnref{P7}, we obtain
\begin{equation} \eqnlabel{P13}
I(W_2; \Yv|\Xv_1)= nI(V; Y|U, X_{1},Q)+ \Sigma_1-\Sigma_2.
\end{equation}
Similarly, using \eqnref{P10} in \eqnref{P8}
\begin{equation} \eqnlabel{P14}
I(W_2; \Yv_1|\Xv_1) = nI(V; Y_{1}|U, X_{1},Q)+ {\hat \Sigma_1}-{\hat \Sigma_2}.
\end{equation}
Substituting \eqnref{P13} and \eqnref{P14} in \eqnref{equivocation} results in
\begin{align} \eqnlabel{P15}
\frac{1}{n}&H(W_2|\Xv_1,\Yv_1) \nonumber \\
 &\le I(V;Y |U,X_1,Q)- I(V; Y_{1}|U, X_{1},Q)+ \delta_n.
\end{align}
Using \eqnref{P17} in \eqnref{P15}, we obtain the desired the bound \eqnref{R2QOUTER} on rate
$R_2$.

We next prove the bound on the sum rate \eqnref{upperboundQ}.
\begin{equation}
\begin{split}\label{eq:sum1}
n(R_1+R_2)
&=   I(W_1,W_2;{\bf Y})+H(W_1,W_2|{\bf Y})\\
&\le  I({\bf X}_1,W_2; {\bf Y})+ n \delta_n \\
&\le I({\bf X}_1,W_2; {\bf Y})-[H(W_2)\\
& \hspace{0.2in} - H(W_2|{\bf X}_1,{\bf Y}_1) -n\epsilon]+ n\delta_n\\
& = I({\bf X}_1; {\bf Y})+ I(W_2;{\bf Y}|{\bf X}_1) \\
& \hspace{0.2in} - I(W_2;{\bf Y}_1|{\bf X}_1) + n(\delta_n+\epsilon)
\end{split}
\end{equation}
where the second inequality follows from the perfect secrecy \eqnref{Hlim2}. Using \eqnref{P7},
\eqnref{P8} and Lemma~\ref{lem1}, we have
\begin{align}
I(&W_2; {\Yv}|{\bf X}_1)-I(W_2; {\bf Y}_{1}|{\bf
X}_1) \nonumber \\
&=\sum_{i=1}^{n}\bigl[I(W_2;Y_i|U_i, X_{1,i})-I(W_2;Y_{1,i}|U_i, X_{1,i})\bigr]
\end{align}
Hence, (\ref{eq:sum1}) can be rewritten as
\begin{align*}
&~n(R_1+R_2) \nonumber\\
&\le  \sum_{i=1}^{n} \bigl[ I(\Xv_1; Y_i|\Yv^{i-1})+I(W_2;Y_i|U_i, X_{1,i})  \nonumber \\
& \hspace{0.2in} -I(W_2;Y_{1,i}|U_i, X_{1,i})\bigr] +n(\delta_n+\epsilon) \nonumber \\
&\le \sum_{i=1}^{n}\bigl[ I(\Xv_1, {\bf Y}^{i-1},\tilde{\bf Y}^{i+1}_1;Y_i)+I(W_2;Y_i|U_i, X_{1,i}) \nonumber \\
& \hspace{0.2in}  -I(W_2;Y_{1,i}|U_i, X_{1,i})\bigr] +n(\delta_n+\epsilon) \nonumber \\
&= \sum_{i=1}^{n} \bigl[I(V_i,X_{1,i}; Y_i)-I(V_i;Y_{1,i}|U_i,X_{1,i})\bigr]
+ n(\delta_n+\epsilon) \nonumber\\
&\le  \sum_{i=1}^{n} \bigl[I(V_i,U_i,X_{1,i}; Y_i)-I(V_i;Y_{1,i}|U_i,X_{1,i})\bigr] +
n(\delta_n+\epsilon) 
\end{align*}
where $U_i$ and $V_i$ are defined in \eqnref{U} and \eqnref{V}. Using the same time-sharing
variable approach as before we obtain the sum rate bound \eqnref{upperboundQ}. Moreover, the
Markovity $X_{1,i}-U_i-V_i$ can easily be verified.
\end{proof}

\section{Achievability}

\begin{proof}  {\it (Theorem~\ref{maintheorem})}

Fix $p(u)$, $p(x_1|u)$, $p(v|u)$ and $p(x_2|v)$. Let
\begin{equation} \eqnlabel{R3}
R_3=R_2+I(V;Y_1|X_1,U).
\end{equation}

{\bf Codebook generation:} Generate a random typical sequence $\uv$, with probability $p(\uv)=
\prod_{i=1}^{n}p(u_i).$ We assume that both transmitters and the common receiver know the sequence
$\uv$.

Generate $M_1=2^{nR_1}$ sequences $\xv_1$, each with probability $p(\xv_1|\uv)=
\prod_{i=1}^{n}p(x_{1,i}|u_i)$. Label them $\xv_1(w_1)$, $w_1\in\{1,\dots,M_1\}$.

Generate $M_3=2^{nR_3}$ sequences $\vv$ with probability $p(\vv|\uv)= \prod_{i=1}^{n}p(v_i|u_i)$.
Label them $\vv(w_2,l)$, $w_2 \in \{1, \dots,2^{nR_2}\}$, $l \in \{1, \dots, 2^{nI(V;Y_1|X_1,U)} \}$.

{\bf Encoding:} To send message $w_1\in\mathcal W_1$, user $1$ sends codeword $\xv_1(w_1)$. To send
message $w_2 \in\mathcal W_2$, user $2$ uses stochastic encoder $f$, and encoder $2$ uniformly
randomly chooses an codeword $\vv(w_2,l)$. That is, the encoder chooses randomly a codeword $\vv(w_2,l)$ from a bin $w_2$.
Finally, user $2$ generates the channel input sequences $\xv_2$ according to $p(x_2|v)$.

{\bf Decoding:} Let $\A$ denote the set of typical $(\uv,\xv_1,\vv,\yv)$ sequences. Decoder
chooses the pair $(w_1,w_2)$ such that $(\uv,\xv_1(w_1),\vv(w_2,l),\yv)\in\A$ if such a pair $(w_1,w_2)$ exists and is unique; otherwise, an error is declared.

{\bf Probability of error:} Define the events
\begin{align}
E_{w_1,w_2}=\{ (\uv, \xv_1(w_1),\vv(w_2,l),\yv )    \in\A \}.
\end{align}
Without loss of generality, we can assume that $(w_1,w_2)=(1,1)$ was sent. From the union bound, the
error probability is given by
\begin{align}
P_e\le& P\{E_{1,1}^{c}|(1,1)\}+\sum_{w_1\neq 1}P\{E_{w_1,1}|(1,1)\} \nonumber \\
&+\sum_{w_2\neq 1} \sum_l  P\{E_{1,w_2}|(1,1)\} \nonumber\\
&+\sum_{w_1\neq 1} \sum_{w_2\neq 1} \sum_l P\{E_{w_1,w_2}|(1,1)\} \label{eq:pe0}
\end{align}
From the AEP and~\cite[Thm. 14.2.1, 14.2.3]{Cover}, it follows that
\begin{align}
P\{E_{1,1}^{c}|(1,1)\}&\le \delta\\
P\{E_{w_1,1}|(1,1)\}&\le 2^{-n[I(X_1;Y|V,U)-\delta]}\\
P\{E_{1,w_2}|(1,1)\}&\le 2^{-n[I(V;Y|X_1,U)-\delta]}\\
P\{E_{w_1,w_2}|(1,1)\}&\le 2^{-n[I(X_1,V;Y|U)-\delta]}
\end{align}
where $\delta\rightarrow 0$ as $n\rightarrow \infty$. Hence, (\ref{eq:pe0}) is bounded by
\begin{align}
P_e& \le
\delta+2^{nR_1}2^{-n(I(X_1;Y|V,U)-\delta)}+2^{nR_3}2^{-n(I(V;Y|X_1,U)-\delta)}\nonumber\\
&+2^{n(R_1+R_3)}2^{-n(I(X_1,V;Y|U)-\delta)} \label{eq:pe}
\end{align}
implying that we must choose
\begin{align}
R_1& \le I(X_1; Y|V,U)\\
R_3& \le I(V; Y|X_1,U)\\
R_1+R_3& \le I(X_1,V; Y|U)
\end{align}
to guarantee $P_e\rightarrow 0$ as $n$ gets large.

{\bf Equivocation:} We consider the normalized equivocation.
\begin{align}
 H(W_2&|\Yv_1,\Xv_1) \nonumber \\
&\ge H(W_2|\Yv_1,\Xv_1,\Uv)\nonumber\\
&= H(W_2,\Yv_1|\Xv_1,\Uv)-H(\Yv_1|\Xv_1,\Uv)\nonumber\\
&=H(W_2,\Yv_1,\Vv|\Xv_1,\Uv)-H(\Vv|W_2,\Yv_1,\Xv_1,\Uv) \nonumber\\
& \hspace{0.2in}-H(\Yv_1|\Xv_1,\Uv) \nonumber\\
&=H(W_2,\Vv|\Xv_1,\Uv)+H(\Yv_1|W_2,\Vv,\Xv_1,\Uv) \nonumber \\
& \hspace{0.2in} -H(\Vv|W_2,\Yv_1,\Xv_1,\Uv)-H(\Yv_1|\Xv_1,\Uv)\nonumber\\
&\ge H(\Vv|\Xv_1,\Uv)+H(\Yv_1|\Vv,\Xv_1,\Uv)\nonumber \\
& \hspace{0.2in}-H(\Vv|W_2,\Yv_1,\Xv_1,\Uv) -H(\Yv_1|\Xv_1,\Uv)\nonumber\\
&=H(\Vv|\Xv_1,\Uv)-H(\Vv|W_2,\Yv_1,\Xv_1,\Uv)\nonumber \\
& \hspace{0.2in}-I(\Vv;\Yv_1|\Xv_1,\Uv)\label{eq:ev0}
\end{align}

The first term in (\ref{eq:ev0}) is given by
\begin{align}
H(\Vv|\Xv_1,\Uv)=H(\Vv|\Uv)=nR_3\label{eq:ev1}
\end{align}
where the first equality follows from the Markov chain $\Vv-\Uv-\Xv_1$, and the second equality
because given $\Uv=\uv$, $\Vv$ has $2^{nR_3}$ possible values with equal probability.

We next show that $H(\Vv|W_2,\Yv_1,\Xv_1,\Uv)\le n\delta_1$, where $\delta_1 \rightarrow 0$ as
$n\rightarrow \infty$. Let $W_2=w_2$. User $2$ then sends a codeword $\vv(w_2,l)$. Let $\lambda_{w_2}$ denote the average probability of error that
user $1$ does not decode  $\vv(w_2,l)$ correctly given the information $W_2=w_2$. Following the
joint typical decoding approach, we have $\lambda_{w_2} \rightarrow 0$ as $n\rightarrow \infty$.
Therefore, Fano's inequality implies that
\begin{align*}
H(\Vv|W_2=w_2,\Yv_1,\Xv_1,\Uv)\le 1+\lambda_{w_2}(nR_3-nR_2) \triangleq n\delta_1.
\end{align*}
Hence
\begin{align}
H(\Vv&|W_2,\Yv_1,\Xv_1,\Uv)= \nonumber \\
& \sum_{w_2\in \mathcal W_2}p(W_2=w_2) H(\Vv|W_2=w_2,\Yv_1,\Xv_1,\Uv) \le n\delta_1. \label{eq:ev2}
\end{align}

Finally, the third term in (\ref{eq:ev0}) can be bounded by
\begin{equation} \eqnlabel{eq:ev3}
I(\Vv; \Yv_1| \Xv_1, \Uv) \le nI(V;Y_1|X_1,U) + n\delta_2
\end{equation}
where $\delta_2 \rightarrow 0$ as $n \rightarrow \infty$. The proof follows the proof in
\cite[Lemma $8$]{Wyner75}.

Therefore, by using \eqnref{R3}, (\ref{eq:ev1}), (\ref{eq:ev2}), and \eqnref{eq:ev3}, we can
rewrite (\ref{eq:ev0}) as
\begin{align}
H(W_2|\Xv_1,\Yv_1) &\ge n R_3-nI(V;Y_1|X_1,U) -n(\delta_1+\delta_2) \nonumber\\
&=n R_2-n\epsilon
\end{align}
where $\epsilon \triangleq \delta_1+\delta_2$.
\end{proof}

\section{Discussion and Implications}
To show the impact of secret communication on the achievable rates in MACC, we present two examples: the half-duplex MACC and the Gaussian MACC. To
simplify calculations, we consider the following corollary which gives a weaker inner bound used
in the rest of the paper.
\begin{MyCorollary} \label{cor:ach}
The rates in the closure of the convex hull of all $(R_1, R_2)$ satisfying
\begin{align}
R_1& \le I(X_1; Y|X_2) \label{lb1}\\
R_2& \le I(X_2; Y|X_1)-I(X_2;Y_1|X_1)  \label{lb2}\\
R_1+R_2& \le I(X_1,X_2; Y)-I(X_2;Y_1|X_1) \label{lb3}
\end{align}
for fixed product distribution $p(x_1)p(x_2)$ on ${\Xc_1}\times{\Xc_2}$ is achievable in MACC.
\newline
\begin{proof}
Corollary follows by choosing $V=X_2$ and $U$ independent from $X_1$ and $X_2$ in
Theorem~\ref{maintheorem}.
\end{proof}
\end{MyCorollary}

Binary inputs are to be communicated from the both users under a half-duplex model in which user
$1$ cannot listen and transmit at the same time. Therefore $X_2 \in \{0,1\}$ and $X_1 \in
\{\varnothing, 0,1 \}$. Null symbol $\varnothing$ models the listening period of user $1$. When
$X_1=\varnothing$, user $1$ observes the output $Y_1=Y$; when user $1$ transmits,  $X_1 \in \{0,1
\}$, the output $Y_1$ is the null symbol, no matter what user $2$ sends. When both users transmit,
the MAC channel to the destination is given by the $\text{mod }2$ sum $Y=X_1 \oplus X_2$.
Otherwise, $Y=X_2$. In summary,
\begin{alignat}{3}
Y=X_1 \oplus X_2,&\quad  Y_1=\varnothing, & \qquad \mbox{if } X_1 \neq \varnothing &\\
Y=X_2, & \quad Y_1=Y, &    \qquad \mbox{if } X_1=\varnothing
\end{alignat}
Denote $P=P[X_1=1]$ and $D=P[X_1=\varnothing]$.
Rates (\ref{lb1})-(\ref{lb3}) for this channel can be shown to be
\begin{align}
R_1& \le h(P) \label{r1}\\
R_2& \le H(X_2) (1-D)  \label{r2}\\
R_1+R_2& \le H(Y) - H(X_2)D. \label{r3}
\end{align}
If we assume the inputs at user $2$ are equally likely, then $H(Y)=1$.
The rates (\ref{r1})-(\ref{r3}) become
\begin{align}
R_1& \le h(P) \label{r11}\\
R_2& \le 1-D  \label{r22}\\
R_1+R_2& \le 1 - D \label{r33}
\end{align}
and the secrecy constraint (\ref{r22}) becomes irrelevant. The achievable rates are determined by the amount of time user $1$ listens: the more user $1$ listens, the more user $2$ must equivocate rather than communicate. The best strategy  is then for user $1$ to transmit all the time $(D=0)$, thus achieving the full capacity region of the conventional MAC.

In the other limiting case in which user $1$ only listens ($D=1$), user $2$ cannot send
information because user $1$ hears it ($Y_1=X_2$). In fact,  the channel reduces to the special
case of the channel considered in \cite{CsiszarKorner78} and the conclusion is agreeable with that
of \cite{CsiszarKorner78}.  In the example, the fact that $R_2=0$ is due to the very special
channel $Y_1=Y$. In the more general case in which $Y_1$ is a noiser observation of $X_2$ than
$Y$, user 2 can still ``squeeze'' some information through even if user $1$  listens all the time.
Nonetheless, this example illustrates the fundamental behavior in the MACC, that can be observed
from Corollary~\ref{cor:ach}, Eq. (\ref{lb2}): the more user $1$ decides to listen, the more user
$2$ has to equivocate and his achievable rate is lower.

We next consider the Gaussian channel
\begin{align}
Y =  &X_1 +X_2 + Z\\
Y_1 =& X_2+Z_1
\end{align}
where $Z$ and $Z_1$ are independent zero-mean Gaussian random variables with variance $N$ and $N_1$, respectively.
The code definition is the same as given in Section~\ref{s: Channel} with the addition of the power constraints
\begin{equation} \eqnlabel{powerconstraint}
\frac{1}{n} \sum_{i=1}^n E[X_{ti}^2] \le P_t, \qquad t=1,2.
\end{equation}
\begin{MyCorollary}
The rates in the closure of the convex hull of all $(R_1,
R_2)$ satisfying
\begin{align}
R_1 & \le C \left( \frac{ P_1}{N} \right )  \eqnlabel{1} \\
R_2&  \le C \left( \frac{ P_2}{N} \right ) - C \left( \frac{ P_2}{N_1} \right )  \eqnlabel{2} \\
R_1+R_2& \le C \left( \frac{P_1+ P_2}{N} \right )- C \left( \frac{ P_2}{N_1} \right ) . \eqnlabel{3}
\end{align}
\end{MyCorollary}
Corollary follows from Theorem~\ref{maintheorem} by independently  choosing $X_t \sim \Nc [0,P_t]$ for $t=1,2$.
\subsection*{Future Work}
It is conceivable that the outer bounds given in Theorem~\ref{upperbound} can be strengthened to coincide with the lower bounds of Theorem~\ref{maintheorem}. Investigating this possibility and determining the MACC capacity are the subjects of our future work. Moreover, the formulation of this problem in which the objective is to maximize rates under the secrecy constraint follows the definition of Wyner \cite{Wyner75}. However, different objectives can be envisioned, in which user $1$ is more interested in eavesdroping than in maximizing its rate.  It would be interesting to compare the conclusions that follow from the two problem formulations.

\bibliographystyle{IEEEtran}
\bibliography{referencesCOOP}

\end{document}